# A Controllable Perceptual Feature Generative Model for Melody Harmonization via Conditional Variational Autoencoder

**Dengyun Huang[1], Yonghua Zhu[1]\***

[1]Shanghai Film Academy, Shanghai University, Shanghai, China

{hdy, zyh}@shu.edu.cn

**Abstract**

While Large Language Models (LLMs) make symbolic music generation increasingly accessible, producing music with distinctive composition and rich expressiveness remains a significant challenge. Many studies have introduced emotion models to guide the generative process. However, these approaches still fall short of delivering novelty and creativity. In the field of Music Information Retrieval (MIR), auditory perception is recognized as a key dimension of musical experience, offering insights into both compositional intent and emotional patterns. To this end, we propose a neural network named CPFG-Net, along with a transformation algorithm that maps perceptual feature values to chord representations, enabling melody harmonization. The system can controllably predict sequences of perceptual features and tonal structures from given melodies, and subsequently generate harmonically coherent chord progressions. Our network is trained on our newly constructed perceptual feature dataset BCPT-220K, derived from classical music. Experimental results show state-of-the-art perceptual feature prediction capability of our model as well as demonstrate our musical expressiveness and creativity in chord inference. This work offers a novel perspective on melody harmonization and contributes to broader music generation tasks. Our symbolic-based model can be easily extended to audio-based models.

**Spiral Visualization** —

http://music.designerly.top/spiral_visualization/

**Project Code** —

https://github.com/huang0251/CPFG_model

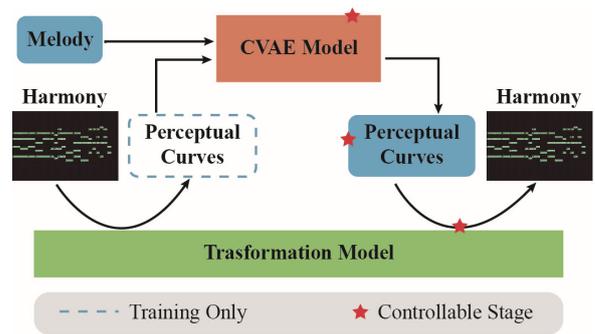

Figure 1: Simple pipeline of our approach. Within this workflow, features can be controlled at three stages: during the sampling from the latent space, during the modification of the generated feature curves, and during the transformation of the feature curves into target harmonies.

# Introduction

Melody harmonization is a conditional music generation task that refers to generating harmonic accompaniments for the given melody, where chord generation is a crucial goal (Ji et al. 2023). Like other tasks in music generation, the generated harmony is often seen as lacking artistic depth, and expressiveness. For these challenges, several recent studies have increasingly focused on incorporating various controllable factors into deep learning network models, such as musical style (Yi et al. 2022; Zhu et al. 2020), harmonicity (Wu et al. 2023), texture (Min et al. 2023), and emotion or sentiment (Ji and Yang 2023; Huang et al. 2024; Hahn et al. 2024). While these factors contribute to expressiveness and creativity (Carnovalini and Rodà 2020), their methods mainly guide the generated content via predefined coarse-grained feature targets, which may limit their ability of precise control of chord inference. Raposo and Soares (2025) apply controlled deviations in a probabilistic framework to keep creative exploration. Although this enhances novelty of chord inference, it compromises the connection with sentiment factors which contribute to emotional expressiveness.

To mitigate these problems, some notable studies (Guo et al. 2022; Verstraelen 2019; Ruiz-Marcos 2022; Wilson et al. 2023) have employed Musical Tonal Tension (MTT) (Krumhansl 1996; Lerdahl and Krumhansl 2007), which plays a crucial role in auditory sensory cognition associated with mood (Sun et al. 2020), as the core parameter for guiding the generation of creative music. However, although fine-grained MTT features reveal musical expression and expectation (Navarro-Cáceres et al. 2020), the end-to-end framework in these workflows may result in limited transparency and interaction. In addition, note-level encoding and decoding in these works may be limited in capturing the dynamics of MTT. On the other hand, no study has applied controllable MTT features specifically to the task of melody harmonization, and most chord generation tasks often take into account a finite set of chord types (Ji et al. 2023), potentially limiting creative exploration of chord inference.

To address the above challenges, in this paper, we propose a modeling framework centered on MTT for generating or refining harmonic accompaniments to the melody. Firstly, to analyze the quantifiable MTT in any chord progression more accurately, we establish a mathematical model that facilitates the bidirectional transformation between perceptual MTT features and harmonic content. Leveraging this model, which is robust in application to the majority of music genres, we construct a novel classical music perceptual dataset based on the Bach chorale corpus. Subsequently, we design a lightweight perceptual feature generative network to capture the patterns of perceptual MTT features. Our strategy is more controllable and transparent, and enables the generation of expressive and creative chord progressions. An overview of our pipeline is shown in Figure 1.

Our main contributions can be summarized as follows:

- To the best of our knowledge, we are the first to explore melody harmonization driven by neural network-predicted perceptual features. Our strategy is compatible with any chord categories, supporting multi-stage control.
- We introduce an accurate transformation model between perceptual features and harmony, which is compatible with key modulation and tonicization scenarios.
- We construct a controllable perceptual feature generation network CPFG-Net, which can output diverse perceptual tension features.
- We offer a new perceptual feature dataset named BCPT-220K, which is derived from classical music.
- Objective experiments demonstrate our method's best performance in reconstruction or generation of perceptual features. Generation test shows the creativity, controllability, and expressiveness of diverse harmonization.

## Related Works

### AI-AMG

The growing influence of AI-based Affective Music Generation (AI-AMG) (Dash and Agres 2024) has become increasingly evident. Tan and Herremans (2020) learn controllable emotion representations by modeling their corresponding quantifiable low-level attributes. Hung et al. (2021) present a dataset named EMOPIA for symbolic music. Zhang et al. (2021; 2023) contribute to emotion recognition and classification. Then, some well-known approaches (Sulun et al. 2022; Ji and Yang 2023; Huang et al. 2024; Hahn et al. 2024) attempt to improve affective expressiveness and musical novelty by incorporating emotion, aiming to control the sentiment in the generated music to influence thematic development. However, coarse-grained emotion remains insufficient for fine-grained control over musical texture. This suggests the value of shifting the research focus one step prior to emotional emergence—auditory sensory cognition.

### MTT in Perceptual Features

Several well-known theories, such as Spiral Array (Chew 2000; Chew 2014), Tonal Pitch Space (TPS) (Lerdahl 1996; Lerdahl 2001) and Tonal Interval Space (TIS) (Bernardes et al. 2016), have proven effective in modeling MTT, aligning well with statistical results from psychoacoustic studies. Recently, TenseMusic (Barchet et al. 2024) uses a linear regression method to predict a single metric of MTT, proving the predictability of MTT. Applying TPS, AuToTen was developed (Ruiz-Marcos 2022; Ruiz-Marcos et al. 2020) for generating music with matched tension profiles. Applying TIS, Navarro-Cáceres et al. (2020) proposed a model for computing tension features of chord progressions. Based on Spiral Array, several tension-aware music generation works (Herremans and Chew 2017; Wilson et al. 2023; Guo 2023; Kang et al. 2024) emerged. Our method is based on Spiral Array due to its sufficient adaptability to various chords, and simplicity of construction.

In chord progression analysis, tension features are typically characterized along three primary dimensions: chord consonance, the distance between adjacent chords (Rocher et al. 2010), and chord strain within a given tonality (Yoo and Lee 2006), all of which are also referenced in our research. Herremans and Chew (2016) introduced computational measures for these features and developed the Midi Miner (Guo et al. 2019) for auto computing, based on Spiral Array. However, their approach exhibits inaccuracies in certain scenarios (see Appendix I for supporting evidence), which are effectively addressed by our proposed model.

### Spiral Array

The Spiral Array 3D space model is composed of several fundamental elements: pitch coordinates set $\mathbb{P}$, chord coordinates set $\mathbb{C}$, and key coordinates set $\mathbb{K}$. The formulas for computing these coordinates are as follows:

$$P(k) = \begin{bmatrix} r\,sin(k\pi/2) \\ rcos(k\pi/2) \\ kh \end{bmatrix}, \quad (1)$$

$$C_{Major}(k) = w_1 \cdot P(k) + w_2 \cdot P(k+1) + w_3 \cdot P(k+4)$$
$$C_{minor}(k) = u_1 \cdot P(k) + u_2 \cdot P(k+1) + u_3 \cdot P(k-3), \quad (2)$$

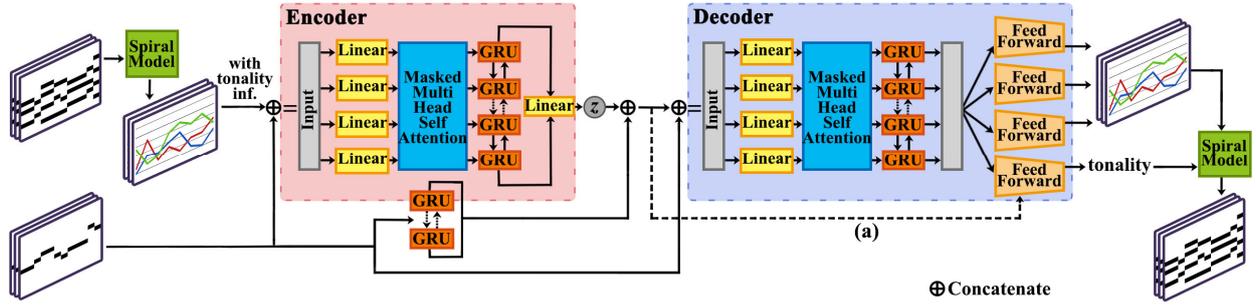

Figure 2: Architecture of perceptual feature reconstruction model for melody harmonization. Dropout and ReLU non-linear activation functions are applied at appropriate locations in the network, which are omitted from the figure for clarity. (a) Concatenated $z$ and melody embedding are imported in tonality feed forward as a skip connection.

$$K_{Major}(k) = \omega_1 \cdot C_M(k) + \omega_2 \cdot C_M(k+1) + \omega_3 \cdot C_M(k-1)$$
$$K_{minor}(k) = v_1 \cdot C_m(k)$$
$$+ v_2 \cdot [\tau_1 \cdot C_M(k+1) + (1-\tau_1) \cdot C_m(k+1)]$$
$$+ v_3 \cdot [\tau_2 \cdot C_M(k-1) + (1-\tau_2) \cdot C_m(k-1)] \quad (3)$$

Here, $r$ is the radius of space model; $h$ is the altitude intercept between adjacent $P$; and $k$ is the index of pitch classes in the circle of fifths, which is generally set to 0 for pitch class 'C'. The parameters $\tau_1$ and $\tau_2$ denote the variations among the natural minor, harmonic minor, and melodic minor scales, and are generally assigned a value of 0.75. Additionally, the three-dimensional weights $w$, $u$, $\omega$ and $v$ satisfy $w = u = \omega = v = [0.5353, 0.2743, 0.1904]$.

## Methods

### Overview

The architectural framework of our proposed method for melody harmonization is illustrated in Figure 2. After training, it can diversely generate harmonies from the input melody and sampled latent representation $z$, with multiple control interfaces. Firstly, we establish a transformation model between perceptual tension metrics and chord progression. Then, we train an improved CVAE (Sohn et al. 2015) model that learns deeper perceptual motivations, where melody sequences are treated as a condition. All methods here can be found in the Code provided following **Abstract**.

### Mathematical Transformation Model

We develop our model based on Spiral Array to obtain the perceptual feature sequences of chord progression $c = [c_1, c_2, ..., c_T]$, where we denote $r = 1$ and $h = 0.4$. A visualization of our model is shown at Spiral Visualization link after paper abstract.

---

Algorithm 1: Transform pitches into Spiral 3D space.
**Input**: A chord sequence: $c = [c_1, c_2, ..., c_T]$; where $c_t = [p_{t,1}, p_{t,2}, ..., p_{t,N_t}]$, $1 \le t \le T$.
**Parameter**: Variable $k \in [-11, 11]$; $\ell$ denotes label of pitch coordinate $P$; and $p$ denotes pitch.
**Output**: A sequence of index $k$: $\mathbb{k} = [\mathbb{k}_1, \mathbb{k}_2, ..., \mathbb{k}_T]$; where $\mathbb{k}_t = [k_{t,1}, k_{t,2}, ..., k_{t,N_t}]$, $1 \le t \le T$.
**for** $t = 1, 2, ..., T$ **do**
  **for** $n = 1, 2, ..., N_t$ **do**
    $\mathbb{k}_{t,n} \leftarrow \{k \mid P_k \in \mathbb{P}, \ell_k = p_{t,n}\}$.
  **end**
  There exist sets $\mathbb{k}_{t,1}, \mathbb{k}_{t,2}, ..., \mathbb{k}_{t,N_t}$
  $\mathbf{k}_t \leftarrow \{\mathbb{k}_t \mid \mathbb{k}_t \in \operatorname*{argmin}_{k_{t,n} \in \mathbb{k}_{t,n}} f(k_{t,1}, ..., k_{t,N_t})\}$
  where $f(k_1, ..., k_N) := \max_{1 \le i < j \le N} \|P_{k_i} - P_{k_j}\|$
  and $size(\mathbb{k}_t) == N_t$
**end**
Calculate $\mathbb{k} \leftarrow \operatorname*{argmin}_{\mathbb{k}_t \in \mathbf{k}_t} \sum_{t=2}^{T} \|ce(\mathbb{k}_t) - ce(\mathbb{k}_{t-1})\|$
where $ce(k_1, ..., k_N) := \sum_{i=1}^{N} P_{k_i} \times \frac{1}{N}$
**return** $\mathbb{k}$

---

### Calculating Perceptual Features from Chords

In our strategy, enharmonic keys are excluded, and all pitches must adopt consistent representation labels (e.g., no simultaneity of 'A$^b$' and 'G$^\#$'). Repeated pitches with the same label should only be kept once. Furthermore, the order of notes does not change the type of chords. Then, the $\mathbb{P}$ within the Spiral Array is transformed from the sequence shown in Table 1.

| Pitch | G | D | A | E | B | G$^b$ | D$^b$ | A$^b$ | E$^b$ | B$^b$ | F | ↵ |
|---|---|---|---|---|---|---|---|---|---|---|---|---|
| $k$ index | -11 | -10 | -9 | -8 | -7 | -6 | -5 | -4 | -3 | -2 | -1 | ↵ |
| Pitch | C | G | D | A | E | B | G$^b$ | D$^b$ | A$^b$ | E$^b$ | B$^b$ | F |
| $k$ index | 0 | 1 | 2 | 3 | 4 | 5 | 6 | 7 | 8 | 9 | 10 | 11 |

Table 1: Pitch sequence and their indices within the Spiral Array, in our strategy.

Accordingly, given $c = [c_1, c_2, ..., c_T]$, the index $k$ of each pitch in the progression is calculated through Algorithm 1. We use Beam Search (Meister et al. 2020) to deal with the optimization problem when calculating $\mathbb{k}$. Once we obtain the index $k$ of all pitches and confirm their pitch coordinates, we further compute three perceptual feature sequences from $c$: *tension*, *distance*, and *strain*. Specifically, in our strategy, for chord $c$, we can compute $Tension(c)$, $Distance(c_1 \to c_2)$, and $Strain(c)$ as follows:

$$Tension(c) = \max_{1 \le i < j \le N} \|P_i - P_j\|. \tag{4}$$

$$Distance(c_1 \to c_2) = \|ce_{c_1} - ce_{c_2}\|, \tag{5}$$

where
$$ce_c = \sum_{i=1}^{N} P_i \times \frac{1}{N}. \tag{6}$$

$$Strain(c) = \|ce_c - K_{key}\|, \tag{7}$$

where
$$K_{key} = \arg \min_{K \in \mathbb{K}} \|K - ce_{key}\|, \tag{8}$$

and
$$ce_{key} = \sum_{j=1}^{M} P_j \times \frac{1}{M}. \tag{9}$$

Here, $N$ is the number of $P$ in chord $c$. $M$ is the total number of pitches in the music piece for which tonality is to be calculated. Then, for sequence length $T$, the results are: *tension* sequence $t = [t_1, t_2, ..., t_T]$, *distance* sequence $d = [d_1, d_2, ..., d_T]$, and *strain* sequence $s = [s_1, s_2, ..., s_T]$.

**Chords Recovery from Perceptual Features**

We construct a chord library encompassing all possible chord types with note number restricted to five. Following the chord representation method proposed by Jiang et al. (2019), we implement a general classification scheme that enables the chord library to provide pre-selected chord sets for custom style requirements.

Given three perceptual feature sequences $t$, $d$ and $s$, alongside tonality information, we propose the autoregressive reconstruction of a coherent $c = [c_1, c_2, ..., c_T]$, where $c_i \in \mathbb{L}$ for $i = 1, 2, ..., T$, and $\mathbb{L}$ denotes the chord library. Target $c$ is computed as follows:

$$c = \arg \min_{c_i \in \mathbb{L}} \sum_{i=1}^{T} \begin{cases} \alpha_{rec}|\hat{T}(c_i) - \hat{t}_i| + \gamma_{rec}|\hat{S}(c_i) - \hat{s}_i|, & i = 1 \\ \alpha_{rec}|\hat{T}(c_i) - \hat{t}_i| + \beta_{rec}|\hat{D}(c_i \to c_{i-1}) - \hat{d}_i| \\ \quad + \gamma_{rec}|\hat{S}(c_i) - \hat{s}_i|, & i > 1 \end{cases} \tag{10}$$

Here, $\hat{T}(c)$, $\hat{D}(c_1 \to c_2)$, and $\hat{S}(c)$ respectively represent $Tension(c)$, $Distance(c_1 \to c)$, and $Strain(c)$. Meanwhile, $\hat{t}_i$, $\hat{d}_i$, and $\hat{s}_i$ respectively represent *tension* value, *distance* value and *strain* value of the sequence at step $i$. In this **rec**overy process, when $i = 1$, we take $\alpha_{rec} + \gamma_{rec} = 1$, and when $i > 1$, we take $\alpha_{rec} + \beta_{rec} + \gamma_{rec} = 1$. Before this process, $t$, $d$ and $s$ can be manually modified to meet custom requirements.

**CPFG-Net**

As we require output diversity, compatibility with variable length, controllability of perceptual features, and extensibility to key modulation, we follow the theoretical foundations of the Conditional Variational Autoencoder (CVAE) (Sohn et al. 2015), with appropriate modifications. For sequence encoding, we employ Gated Recurrent Units (GRUs) (Chung et al. 2014) and the Attention Mechanism (Vaswani et al. 2017) in the encoder and decoder.

This model, shown in Figure 2, operates over three types of variables: the input variable $x$, the conditional variable $y$, and the latent variable $z$. During training, both the encoder and decoder are employed. The reconstruction targets include $t$, $d$ and $s$ as floating-point types and tonality feature $k$ as a categorical variable with 24 classes (0-11 for C major-B major, and 12-23 for C minor-B minor), so $x \in \mathbb{R}^{B \times T \times 27}$, where $B$ is the batch size and $T$ is the sequence length. Meanwhile, for the conditional melody $y \in \mathbb{R}^{B \times T \times E}$, where $E = 1+128+1+1 = 131$, the feature dimension $E$ is composed of two <PAD> with size 1, one-hot pitch with size 128, and weight token with size 1. During inference, the network samples from $z$ and generates outputs conditioned on $y$. The end of decoder is divided into four separate pipelines, as the four output sequences exhibit distinct data distributions. When decoding output without key modulation, in the tonality pipeline, the result is predicted by a feedforward network: $k = FFN(Cat[Sample(z), GRU(y)]) \in \mathbb{R}^{B \times 24}$, as shown in part (a) of Figure 2.

The principal objective is to minimize the negative conditional Evidence Lower Bound (ELBO) (Kingma and Welling 2013). The optimization target consists of two components: the reconstruction loss and the Kullback–Leibler (KL) divergence loss.

$$\mathcal{L}(\theta, \phi; x, y) = -\mathbb{E}_{z \sim q_\phi(z|x,y)}[\log p_\theta(x|z, y)] \\ + \beta \cdot D_{KL}(q_\phi(z|x,y)||p(z)), \tag{11}$$

Here, the prior $p(z)$ is parameterized as Gaussian distribution $\mathcal{N}(\mu, \sigma^2)$. $\beta$ is a balancing parameter referenced from $\beta$-VAE (Higgins et al. 2017) and uses KL warm-up strategy (Kingma et al. 2016). For the reconstruction loss, we apply Mean Squared Error (MSE) loss to each perceptual feature, and apply Cross-Entropy Loss (CEL) to the one-hot encoded tonality feature. Then, the total loss function is as follows:

$$\mathcal{L}_{total} = \beta \mathcal{L}_{KL} + \lambda_k \mathcal{L}_k + \lambda_t \mathcal{L}_t + \lambda_d \mathcal{L}_d + \lambda_s \mathcal{L}_s, \tag{12}$$

Where the weight hyperparameter $\lambda_k$ is set to 0.1 for loss $\mathcal{L}_k = CEL(k, \hat{k})$ of tonality (*key*) feature, and $\lambda_t, \lambda_d, \lambda_s$ are generally set to 1 for loss $\mathcal{L}_t = MSE(t, \hat{t})$ of *tension* feature, loss $\mathcal{L}_d = MSE(d, \hat{d})$ of *distance* feature, and loss $\mathcal{L}_s = MSE(s, \hat{s})$ of *strain* feature, respectively. The KL loss $\mathcal{L}_{KL} = D_{KL}$ is balanced by $\beta$.

In MusicVAE (Roberts et al. 2017) and similar architectures (Guo et al. 2020; Wang et al. 2020), the dimension of the latent variable typically ranges from 512 to 96. In our case, to achieve a more compact representation while considering the complexity of our data, we initially set the dimension of the latent space to 64.

**Disentanglement Strategy**

We use the disentanglement function (Yang et al. 2019a; Yang et al. 2019b; Higgins et al. 2017) to distinguish the contribution of different dimensions of latent representation to *tension*, *distance*, and *strain*. For an input Sample $S_i$, and its encoded latent representation $z_i \in \mathbb{R}^{64}$, there exist transformation functions $\mathcal{F}$ and $f_\theta$, such that the following formulas hold:

$$Encoder(\mathcal{F}(S_i)) = f_\theta(Encoder(S_i)), \quad (13)$$

$$Decoder(f_\theta(z_i)) = \mathcal{F}(Decoder(z_i)). \quad (14)$$

Moreover, after the network's satisfactory reconstruction, the following formula is expected to hold:

$$\mathcal{F}(S_i) \approx Decoder(f_\theta(Encoder(S_i))). \quad (15)$$

In this way, for samples $\{S_{a,1}, S_{a,2}, \ldots, S_{a,M}\}$ and $\{S_{b,1}, S_{b,2}, \ldots, S_{b,M}\}$ that demonstrate extremely opposite characteristic factor on certain feature between $S_a$ and $S_b$, the primary contributing dimensions can be identified by:

$$z_{diff} = \left| \frac{1}{M}\sum_{i=1}^{M} z_{a,i} - \frac{1}{M}\sum_{i=1}^{M} z_{b,i} \right| \in \mathbb{R}^{64}. \quad (16)$$

Where $M$ denotes sample total number. Then, the first few dimensions with higher values in $z_{diff}$ are considered to have more contribution to this certain feature. When we assume $S_a = \mathcal{F}(S_b)$, the model can reconstruct $S_a$ from $S_b$ using Equation (15), where the function $f_\theta$ is defined as follows:

$$f_\theta(z) = z + \theta \odot z_{diff}, \quad \theta \in \mathbb{R}^{64}. \quad (17)$$

## Experiments and Results Evaluation

**Dataset**

We extract chords and melodies from Johann Sebastian Bach Chorale dataset in Music21 (Cuthbert and Ariza 2010) as our source material, due to its straightforward block chord structure and relatively low annotation error. Following the application of the transformation model, we obtained the corresponding MTT feature sequences. Subsequently, we augment this dataset to comprise approximately $2.2 \times 10^5$ samples. To facilitate disentanglement experiments, each sample is annotated with several feature factors, including, but not limited to, the mean value of each MTT sequence, the Standard Deviation (std) of each sequence, and the Zero-Crossing Rate (ZCR) of gradient of each sequence. The augmentation strategies are detailed in Appendix II. This Bach Chorale Perceptual Tension dataset, named BCPT-220K, is split into a training-test set ratio of 0.8 and 0.2.

**Evaluation Metrics**

**Objective Metrics**

Firstly, to our best knowledge, there is no consensus and objective metrics on assessing the quality of MTT. Therefore, we use MSE to evaluate the reconstruction of MTT. Additionally, we compute the Spearman's Rank Correlation Coefficient (SRCC) (Spearman 1987) to assess the consistency of MTT curve variations. Furthermore, we introduce Mean Recovery Deviation Accumulation (MRDA) to evaluate MTT performance in chord recovery process. The calculation of MRDA is inherited from Equation (10):

$$MRDA = \frac{1}{M}\sum_{j=1}^{M} \frac{1}{T_j}\sum_{i=1}^{T_j}(RD). \quad (18)$$

Where $RD = |\tilde{T}(c_i) - \hat{t}_i| + |\tilde{D}(c_i \to c_{i-1}) - \hat{d}_i| + |\tilde{S}(c_i) - \hat{s}_i|$.

Here, $T_j$ is the chord sequence length of samples $\{S_1, S_2, \ldots, S_M\}$ ($1 \le j \le M$). For tonality labels, we utilize CEL during reconstruction and Accuracy (Acc) during prediction.

Secondly, we assess predicted chord progressions using Chord Coverage (CC), Chord Histogram Entropy (CHE), and Melody-chord Tonal Distance (MCTD) from Yeh et al. (2021). Among these, the CC metric measures the variety of chords in a music piece; the CHE assesses the distribution state of chord occurrences; the MCTD estimates the tonal distance between the melody note and chords. In detail, we calculate the Mean CC which averages CC values across chord counts, due to varying sequence lengths. Notably, we compute MCTD in Spiral Array, through Equation (5). We repeat each experiment 10 times on test set and report the average along with 95% Confidence Interval (CI) for each objective metric.

**Subjective Metrics and Participants**

We conduct a subjective experiment that involves a total of 24 participants (12 males and 12 females). The online survey uses a double-blind strategy so each listener receives audio samples randomly selected from the sample set. To account for the expressiveness of all samples and potential auditory fatigue, each participant first listens to a Noise harmony generated from random MTT values, followed by 10 random samples. The experiment ensures that each sample is listened to by 6 random participants, who evaluate perceptual scores (1–5) for Creativity/Novelty (CN) and Harmony/Consonance (HC).

**Implementation Details**

The experiments are implemented using PyTorch 2.3.0 and the model is trained on an NVIDIA RTX 4090 GPU with 120GB of memory and 24GB VRAM. We adopt the following configuration for our proposed model. A latent space of

| Model | CEL ↓ | MSE | | | SRCC | | |
|---|---|---|---|---|---|---|---|
| | | Tension ↓ | Distance ↓ | Strain ↓ | Tension ↑ | Distance ↑ | Strain ↑ |
| CVAE-G | **0.059**±6.6×10⁻⁴ | 0.016±1.1×10⁻⁴ | 0.026±8.2×10⁻⁵ | 0.027±1.1×10⁻⁴ | 0.892±3.8×10⁻⁵ | 0.975±4.2×10⁻⁵ | 0.965±5.7×10⁻⁵ |
| Transformer | 0.14±8.2×10⁻⁴ | 0.044±3.2×10⁻⁴ | 0.017±8.3×10⁻⁵ | 0.051±1.4×10⁻⁴ | 0.888±4.1×10⁻⁵ | 0.986±5.3×10⁻⁵ | 0.966±6.7×10⁻⁵ |
| CVAE-T | 0.42±7.8×10⁻³ | 0.32±6.9×10⁻³ | 0.20±2.7×10⁻³ | 0.20±2.8×10⁻³ | 0.852±4.1×10⁻⁵ | 0.936±5.9×10⁻⁵ | 0.910±7.2×10⁻⁵ |
| CPFG-Net | 0.078±6×10⁻⁴ | **0.0085**±1.2×10⁻⁴ | **0.0097**±6.1×10⁻⁵ | **0.010**±9.2×10⁻⁵ | **0.893**±3.9×10⁻⁵ | **0.986**±3.6×10⁻⁵ | **0.978**±6.8×10⁻⁵ |

Table 2: MTT and tonality reconstruction performance comparison on test set (values: mean ± 95% CI). **Bold** numbers denote the best values. '↑' or '↓' indicate that larger or smaller value is better.

64 dimensions is employed, with $\beta$ annealed from 0.0 to 1.0. The model is optimized using the Adam, with an initial learning rate of $4\times10^{-4}$ and a gamma value of 0.98. A batch size of 256 is used during training, and models reach a stable state after 50 epochs. Details on the selection of partial parameters, as well as the visualization of PCA (Wold et al. 1987) of latent representation, are provided in Appendix III.

**Comparison with Models**

On the one hand, we evaluate the performance of the generated MTTs. Owing to the lack of networks specifically designed for the reconstruction of MTT, we utilize two state-of-the-art networks with similar implementation for perceptual feature prediction and one additional promising network as baselines. These models are described below:

- CVAE-GRU (CVAE-G): Based on the established VAE model (Guo et al. 2020), we incorporate conditional variables into the encoder and decoder.
- Transformer: A Transformer architecture employed by Guo et al. (2022). In addition, separate FFNs at the end of Decoder are employed (Guo et al. 2020; Dong et al. 2023).
- CVAE-Transformer (CVAE-T): A symbolic musical Transformer-VAE model (Jiang et al. 2020).

Table 2 presents a performance comparison between these models and our proposed model. Except for the CEL metric, our network outperforms all other networks across the remaining objective metrics. Furthermore, we evaluate the MTT quality predicted by three variational networks and compare their MRDA. Considering the time consumption, each experiment randomly selects 1,000 samples from the test set. Table 3 shows the objective results where our model demonstrates lower MRDA. Notably, 'Noise' recovers chords from random MTT values for comparison.

| CVAE-G | CVAE-T | CPFG-Net | Noise |
|---|---|---|---|
| 0.068±0.0022 | 0.071±0.0057 | **0.065**±0.0023 | 0.322±0.013 |

Table 3: MRDA results from chord recovery process of three networks' outputs (values: mean ± 95% CI).

On the other hand, to fairly evaluate the quality of the generated harmonies, we compare the performance of our model with Deepbach (Hadjeres et al. 2017) and Coconet

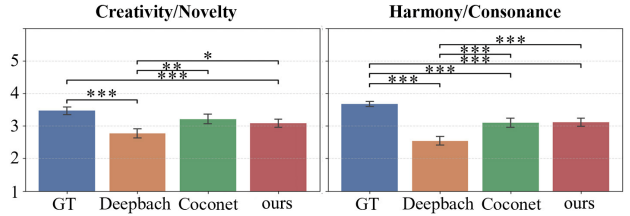

Figure 3: Subjective evaluation results, where GT is the ground truth (*: $p < 0.05$, **: $p < 0.01$, ***: $p < 0.001$).

(Huang et al. 2017), both of which can predict chord progression for a given melody based on Bach Chorale dataset without requiring additional conditions. However, the raw chords predicted by our model are compact and uncomposed, which can lead to unfair comparison. Therefore, during the experiments, all chord progressions are converted into raw form using music21.

| Model | Mean CC | CHE | MCTD |
|---|---|---|---|
| GT | 0.51±0.012 | 2.04±0.051 | 1.36±0.044 |
| Coconet | 0.51±0.021 | 2.18±0.043 | 1.37±0.055 |
| Deepbach | 0.49±0.012 | 2.27±0.038 | 1.48±0.064 |
| ours | 0.57±0.016 | 2.28±0.040 | 1.51±0.054 |
| Noise | 0.67±0.020 | 2.37±0.041 | 1.72±0.025 |

Table 4: Results of objective metrics specifically on assessing melody harmonization.

We randomly collect 10 melody samples from the test set and predict harmonies through three systems: Deepbach, Bach Doodle (Huang et al. 2019) of Coconet, and our model. Table 4 presents the metrics analysis of the generated results. Our model exhibits relatively higher chords richness and lower harmonicity between melody and chords. Subsequently, 40 samples, including both generated harmonies and Ground Truth (GT), are scored by participants across CN and HC metrics, shown in Figure 3. Since each listener may not evaluate results from all systems for the same melody, we conduct significance analysis using the Mann-Whitney U test. The results present that our model is comparable to Coconet in the HC metric, while performing relatively lower in the CN metric.

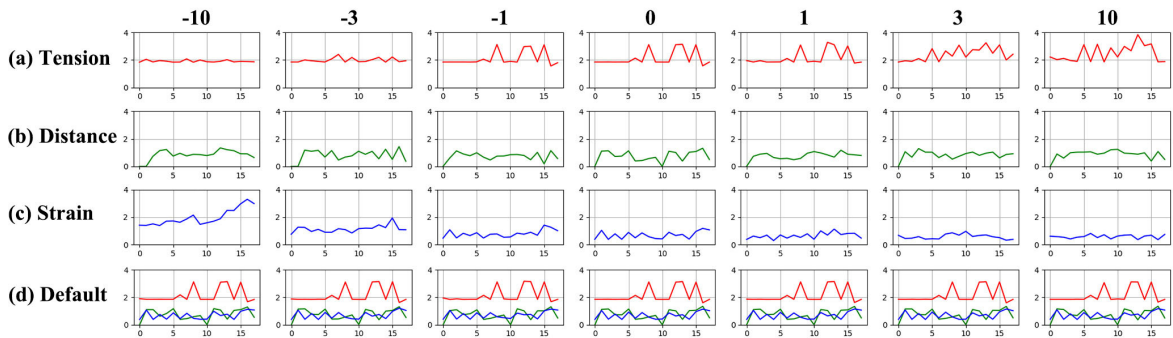

Figure 4: Visualization of perceptual curve variances on a representative sample. (a) Modifying dimensions that contribute the most to std of *tension*. (b) Modifying dimensions that contribute the most to mean of *distance*. (c) Modifying dimensions that contribute the most to ZCR of gradient of *strain*. (d) Modifying dimensions that contribute nothing to any of the features.

| Model | 0 | 1 | 2 | 3 | 4 | 5 | 6 | 7 | 8 | 9 | 10 | 11 | 12 | 13 | 14 | 15 | 16 | 17 | 18 | 19 | 20 | 21 | 22 | 23 |
|---|---|---|---|---|---|---|---|---|---|---|---|---|---|---|---|---|---|---|---|---|---|---|---|---|
| CVAE-G | 17 | - | 2 | 1 | - | 31 | - | 12 | - | - | 19 | - | - | - | 4 | - | - | - | - | 1 | - | 13 | - | - |
| CVAE-T | 5 | 3 | 6 | 5 | 4 | 7 | 4 | 4 | - | 1 | 4 | 5 | 2 | 6 | 2 | 7 | 4 | 7 | 5 | 8 | 3 | - | 4 | 4 |
| w/o TC | 1 | - | 20 | 5 | 2 | 21 | - | 9 | - | 7 | 9 | - | - | - | - | - | 2 | - | - | 3 | - | 20 | - | 1 |
| CPFG-Net | 7 | - | - | - | - | 32 | - | - | - | - | 12 | - | - | - | 34 | - | - | - | - | 8 | - | 7 | - | - |

Table 6: Frequency of tonality occurrence from models' prediction (values in %). In the title line, index *0-11* represents C major-B major, and *12-23* represents C minor-B minor.

## Ablation Study

Further, we conduct an ablation study to evaluate the necessity and contribution of each component in our approach. The main components evaluated are as follows: (i) Tonality Component (TC) shown in part (a) of Figure 2; (ii) Multi-head Self-Attention component (Att1); (iii) GRU cells component (Gs) in both Decoder and Encoder. (iv) An alternative Multi-head Self-Attention (Att2) component replacing GRU. Table 5 shows the objective metrics performance of variant models with partial components, where models without Gs exhibit significant degradation in reconstructing three features, and models without TC achieve lower Acc in tonality prediction. We further evaluate performance of tonality prediction by inputting a typical melody in F major key and generating 1,000 results from the models. The predicted key distributions are shown in Table 6. In our proposed model, tonality predictions no longer exhibit unexpected deviations or noise, resulting in more regularized and consistent outputs. Notably, C major (index 0) and B$^b$ major (index 10) are adjacent to F major (index 5) on the circle of fifths; D minor (index 14) is the relative minor of F major; G minor (index 19) and A minor (index 21) are adjacent to D minor on the circle of fifths.

## Case Study

We subsequently test our model by using the disentanglement strategy. We partially modify samples and get $S_a$ and $S_b$ pairs from BCPT-220K to obtain $z_{diff}$ for certain feature factors. Then, we amplify the latent dimensions using function $f_\theta$ defined in Equation (17). The amplification degrees of $\theta$ are $[-10, -3, -1, 0, 1, 3, 10]$. A visualization of MTT reconstruction variations is illustrated in Figure 4. Subsequently, we reconstruct samples and generate new outputs from the corresponding melodies. The controllable MTT curves enhance creativity in harmony generation. Users can also modify the curves before the chord recovery process, and constrain the chord library $\mathbb{L}$ to enable customize and personalized generation. Due to the challenges of visual explanation of our audio results, we show more details in Appendix IV, where the corresponding music files are provided in the link after **Abstract**.

| TC | Att1 | Gs | Att2 | Acc (%) ↑ | MSE | | |
|---|---|---|---|---|---|---|---|
| | | | | | Tension ↓ | Distance ↓ | Strain ↓ |
| √ | | | | 14.1±0.12 | 1.73 | 1.0 | 0.78 |
| √ | | | √ | 13.9±0.13 | 1.71 | 1.13 | 0.73 |
| | | √ | | 15.1±0.14 | 0.016 | 0.026 | 0.027 |
| | √ | √ | | 14.8±0.13 | 0.0085 | 0.0083 | 0.0096 |
| √ | √ | √ | | **22.8±0.15** | **0.0081** | 0.0092 | 0.01 |

Table 5: Performance of variants from ablation study.

## Conclusions

This paper proposes a novel melody harmonization workflow grounded in MTT features, which is effectively applied to the Bach music dataset, producing chord progressions that are fully expressive. However, the study primarily focuses on the perceptual patterns, without addressing aspects such as chord texture and contrapuntal structure. As a result, the outputs require further refinement and arrangement before practical using. Moreover, due to the inherent nature of the perceptual features, they cannot be fully decoupled, and retain a degree of mutual influence. In future work, we aim to extend this approach to popular music datasets with longer sequence length and key modulation. Additionally, we plan to have adjustable chord density by considering note durations in chord inference.

# Appendix I

In Midi Miner, three features—cloud diameter, cloud momentum, and tensile strain—are modeled based on the Spiral Array, corresponding to the perceptual features of chord dissonance, distance between adjacent chords, and chord strain within a given tonality. The chord consonance value is equivalent to the cloud diameter, which is defined as the maximum diameter of the three-dimensional geometric shape formed by the notes of a chord within the Spiral Array:

$$diameter(C) = \max_{1 \leq i < j \leq N} \left\| P_i(k) - P_j(k) \right\|$$

Here, $P(k)$ is obtained from Equation (1) mentioned in our paper (Related Works Section), representing the coordinates of the pitches in chord $C$, with $N$ being the number of unique pitches in chord $C$. However, determining which $P$ in the Spiral Array a pitch in the chord should be mapped to—i.e., the index $k$—leads to some problems. We discuss some inaccuracies in Midi Miner regarding this, as follows.

Spiral Array arranged the default $P$ set in a sequence from flat to sharp: [......$G^b$, $D^b$, $A^b$, $E^b$, $B^b$, F, C, G, D, A, E, B, $F^\#$, $C^\#$, $G^\#$, $D^\#$, $A^\#$......]. In Midi Miner, Herremans adopted a unified pitch naming strategy and defined the modeling range as: [$G^b$: -6, $D^b$: -5, $A^b$: -4, $E^b$: -3, $B^b$: -2, F: -1, C: 0, G: 1, D: 2, A: 3, E: 4, B: 5], with the C pitch as the center and its $k$ index value set to 0. However, this range loses the cyclic property of the Spiral Array. For example, when the diminished triad [$G^b$, A, C] is input, Midi Miner selects the $k$ indices as $k$ = [-6, 3, 0], omitting the possibility of $k$ = [6, 3, 0]. As a result, the *diameter*([$G^b$, A, C]) yields controversial results. To determine which value is correct, we may add another note, such as E or F, for further evaluation:

- When adding the E pitch, the $k$ = [-6, 3, 0, 4] changes the maximum diameter from 3.8678 to 4.472. However, for $k$ = [6, 3, 0, 4], the maximum diameter remains unchanged compared to $k$ = [6, 3, 0]. Since E does not form a semitone interval with any of the pitches in [$G^b$, A, C], from the perspective of auditory perception and pitch-interval consonance (Krumhansl 2001), the absence of a significant increase in dissonance is more reasonable in this case.
- When adding the F pitch, the maximum diameter for $k$ = [-6, 3, 0, -1] remains the same as that of $k$ = [-6, 3, 0]. However, for $k$ = [6, 3, 0, -1], the maximum diameter increases from 3.1241 to 3.1369, compared to $k$ = [6, 3, 0]. Since F forms a semitone interval with $G^b$, from the perspective of auditory perception and pitch-interval consonance, the increase in dissonance is more reasonable. Considering that diminished triads inherently have high dissonance, such a subtle change also conforms to the phenomenon described by Weber-Fechner Law (Fechner, 1860).

Therefore, we conclude that $k$ = [6, 3, 0] is more appropriate for representing the chord [$G^b$, A, C].

**References**

Krumhansl, C. L. 2001. *Cognitive foundations of musical pitch*. Oxford University Press.

Fechner, G. T. 1860. *Elemente der psychophysik* [Elements of Psychophysics] (Vol. 2). Breitkopf u. Härtel.

## Appendix II

Before calculating MTT features from Bach dataset, we augment the dataset to obtain a more sufficient and balanced dataset. In detail, each metadata sample is expanded using multiple augmentation strategies to generate new variants, including:

- *Modification of data density*: Apart from the melody and corresponding harmony on strong beats, notes and harmonies at other time steps can be randomly removed without violating harmonic inference rules or basic principles of music theory.
- *Melodic pitch alteration*: Notes in the melody may be randomly substituted, with a certain probability, by neighboring pitches in the circle of fifths.
- *Key transposition*: All pitches within the sample can be shifted by a fixed number of steps, thereby altering the tonality of the entire piece. Additionally, transpositions do not cross between major and minor modes.

To avoid imbalances in key distribution, we equalize the number of samples for each of the 12 major keys and each of the 12 minor keys, which can be observed in Figure A3. All augmentation approaches are implemented in the Code provided in Supplementary materials, and are used to generate our BCPT-220K dataset.

To facilitate experimental evaluation and to enrich the high-dimensional feature space of the dataset, each sample is annotated with the following set of labels:

(1) Chorale title: Identifying the original score from which the sample is derived.

(2) Mode: Indicating whether the sample is in a major or minor key.

(3) Standard deviation of each of the three perceptual feature sequences.

(4) Range of each of the three perceptual feature sequences, i.e., maximum minus minimum.

(5) Crossing density of each perceptual feature sequence relative to its mean or median.

(6) Zero-crossing rate of gradient of each feature sequence.

(7) Fast Fourier Transform of each feature sequence.

(8) Mean value of each perceptual feature sequence.

Among these, items 3, 5, 6, and 7 primarily serve to assess the frequency characteristics of the numerical curves, whereas items 4 and 8 reflect information about their magnitude.

## Appendix III

In this Section, we present a series of experiments conducted to determine appropriate model parameter settings and to visualize the structure of the latent space.

**Sorted Variances from PCA**

We initially investigate the impact of various values of the hyperparameter $\beta$, as illustrated in Figure A1, by training the model separately for each setting. We observe that both excessively large and small values of $\beta$ lead to either training instability or deterioration in the quality of generated outputs. Subsequently, we applied Principal Component Analysis (PCA) to the latent variables encoded from 4,096 sam-ples across each model. The corresponding explained variance curves of the principal components, shown in Figure A1, are derived after performing both scale standardization and numerical normalization. These curves reveal that only approximately the first half of the latent dimensions contain significant information. Moreover, models trained with larger $\beta$ values tend to suffer from posterior collapse, resulting in a higher number of inactive or redundant dimensions. To further investigate the effect of latent dimensionality, we reduce the latent space size to 32 and retrain the model. As shown in Figure A1, the slope of curves indicate that the reduced latent space may be insufficient for capturing the complexity of the data.

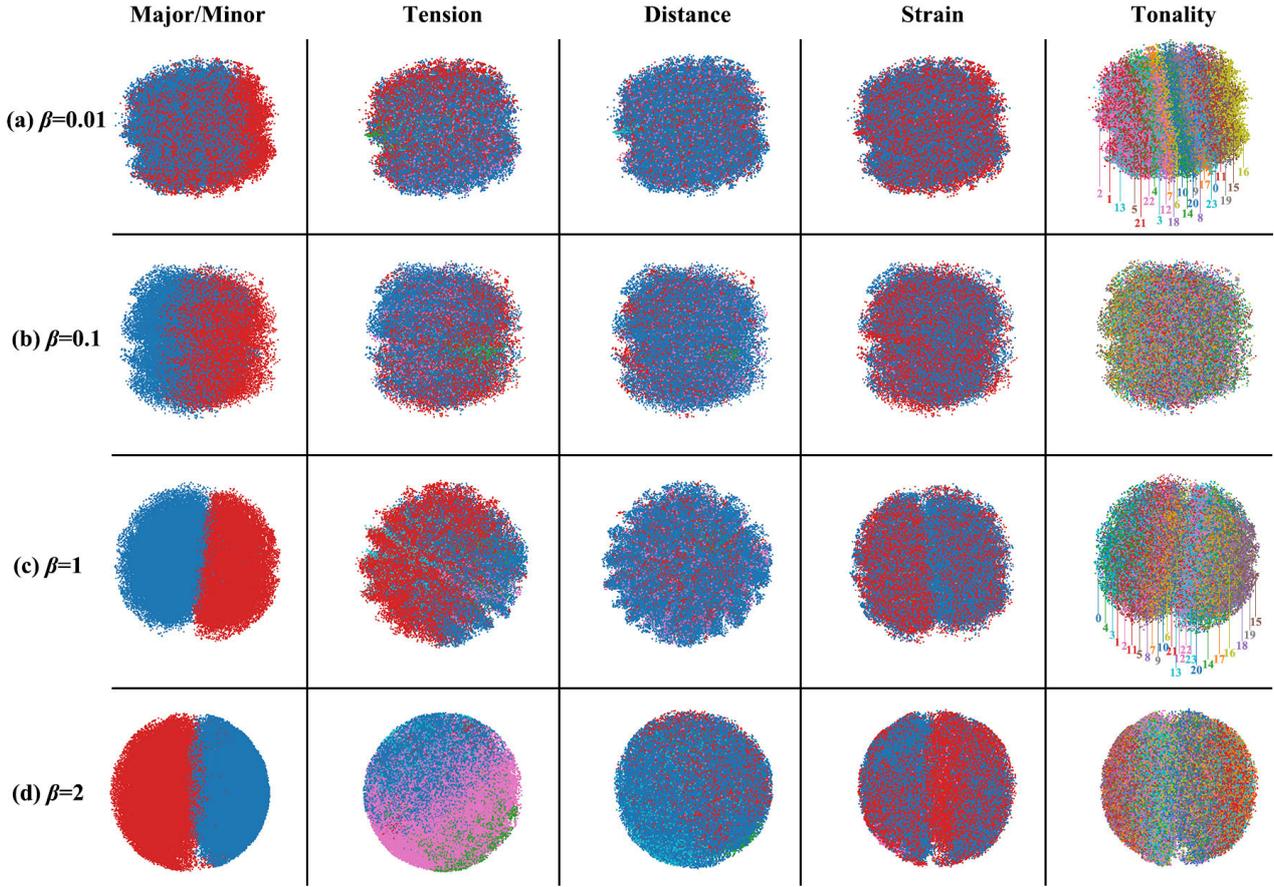

Figure A3: Visualization of latent representations with five important labels, in each model. The labels include: major or minor key, std of *tension* feature sequence of sample, std of *distance*, std of *strain*, and tonality mode of sample. The samples are divided by different std values and expressed in different colors with each quantity.

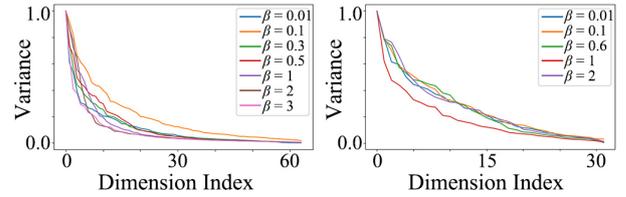

Figure A1: Sorted variance of principal components. Left: 64-dim latent space. Right: 32-dim latent space.

**Dimension Contribution of Latent Variables to Different Features**

To analyze the influence of hyper-parameters on the disentanglement strategy, we further train several models and evaluate their ability to identify key dimensions within the latent variable $z$ that contribute most significantly to specific MTT feature factors. Following the Controlled Variable Methodology, we partially modify samples to get $S_a$ and $S_b$ pairs from BCPT-220K to obtain $z_{diff}$ for certain feature factors. Specifically, since each sample in the dataset is pre-labelled, we initially select the Standard Deviation (std) of the *tension* sequence as the feature factor under analysis. The absolute values of each dimension in $z_{diff}$ are sorted to assess their relative contribution, as visualized in Figure A2. All resulting values are normalized.

| Latent Space | $\beta$ | CEL Loss | KL Loss | MSE Loss | | | Loss | SRCC | | | Mean std |
|---|---|---|---|---|---|---|---|---|---|---|---|
| | | | | Tension | Distance | Strain | | Tension | Distance | Strain | |
| **64-dim** | 0.01 | **0.0044** | 1.41 | **0.0022** | **0.0038** | **0.0045** | **0.025** | 0.885 | 0.986 | 0.980 | 0.663 |
| | 0.1 | 0.0097 | 0.54 | 0.0037 | 0.0069 | 0.0077 | 0.073 | **0.891** | 0.987 | 0.981 | 0.805 |
| | 1 | 0.085 | 0.19 | 0.017 | 0.019 | 0.022 | 0.25 | 0.884 | 0.980 | 0.972 | 0.922 |
| | 2 | 0.17 | **0.13** | 0.028 | 0.026 | 0.026 | 0.36 | 0.885 | 0.978 | 0.970 | **0.942** |
| **32-dim** | 0.01 | 0.0088 | 1.93 | 0.0052 | 0.0081 | 0.0065 | 0.040 | 0.888 | **0.988** | **0.982** | 0.596 |
| | 0.1 | 0.014 | 0.94 | 0.0069 | 0.011 | 0.012 | 0.13 | 0.888 | 0.981 | 0.973 | 0.595 |
| | 1 | 0.17 | 0.57 | 0.035 | 0.043 | 0.046 | 0.69 | 0.887 | 0.964 | 0.952 | 0.620 |
| | 2 | 0.18 | 0.58 | 0.025 | 0.034 | 0.025 | 0.66 | 0.884 | 0.965 | 0.952 | 0.621 |

Table A1: Objective metrics of Models with different hyperparameters. The maximum and minimum values have been coarsely marked. 'std' means standard deviation $\sigma$ from latent space.

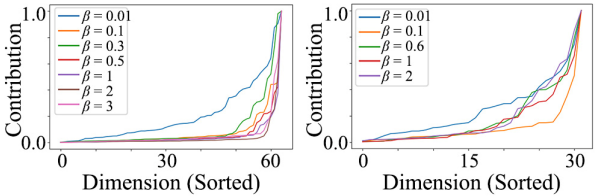

Figure A2: Sorted contribution of dimensions in ascending order to std of *tension* feature. The left and right figure are respectively based on models with 64-dim latent space and 32-dim latent space.

The results from the 64-dimensional latent space indicate that models trained with larger $\beta$ tend to exhibit a higher degree of disentanglement, where fewer dimensions dominate the variation in a given feature factor. In contrast, models using a 32-dimensional latent space demonstrate more entangled representations, with feature variations distributed across a greater number of dimensions.

**Visualization of Latent Representations**
The above experiments suggest that a 64-dimensional latent space is suitable for representing the MTT features. Furthermore, figure A3 shows the 3D PCA visualization of the 64-dimensional latent vectors obtained by encoding all samples using models with different $\beta$ values.

The results indicate that with increasing values of $\beta$, the clustering of samples with similar perceptual features becomes more pronounced. Notably, tonality mode exhibits a strong clustering tendency even at $\beta = 0.01$, and samples with different tonalities form a gradual and continuous transition across the latent space. In Figure A3, rows (a) and (c), provide annotated tonality categories, where indices 0–11 correspond to C major through B major, and indices 12–23 represent C minor through B minor.

**Objective Evaluation**
We evaluate the performance of models under different hyperparameter settings using several objective metrics, shown in Table 2. In addition to the MSE or CEL loss on features reconstruction and KL divergence loss of the VAE, we also assess the Mean std of the latent variable to measure the similarity between the prior distribution $\mathcal{N}(\mathbf{0}, \mathbf{I})$ and the learned posterior distribution $\mathcal{N}(\boldsymbol{\mu}, \boldsymbol{\sigma}^2)$. Besides, we compute the Spearman's Rank Correlation Coefficient (SRCC) between the input and predicted output sequences to evaluate the consistency of their curve variations. Higher Mean-std and SRCC values indicate better model performance in terms of representation quality. All models are trained for 20 epochs. Taking into account the reconstruction loss, the effect of the KL constraint, and the clustering behavior observed in the latent space, we select $\beta = 1$ as our final hyperparameter setting.

**Appendix IV**
As a continuation of the Case Study in our paper, we reconstruct samples and generate new outputs from the melody. Figure A4 illustrates the visualization results. From a visual interpretation perspective, adjacent pitches, i.e., semitones, typically correspond to higher *tension* values, while variations in *distance* and *strain* features are less visually distinguishable. From an auditory perspective, chords in 'Tension +10' are perceived as more dissonant, and the opposite is true in 'Tension -10'; chords in 'Distance +10' and 'Distance -10' sound jumpier between transitions; 'Strain -10' has more off-tone chords (even results in key modulation), while 'Strain -10' introduces more harmonically consistent progressions. The corresponding MIDI files, and their structured, manually arranged versions, along with MP3 audios, are provided in the link after **Abstract**.

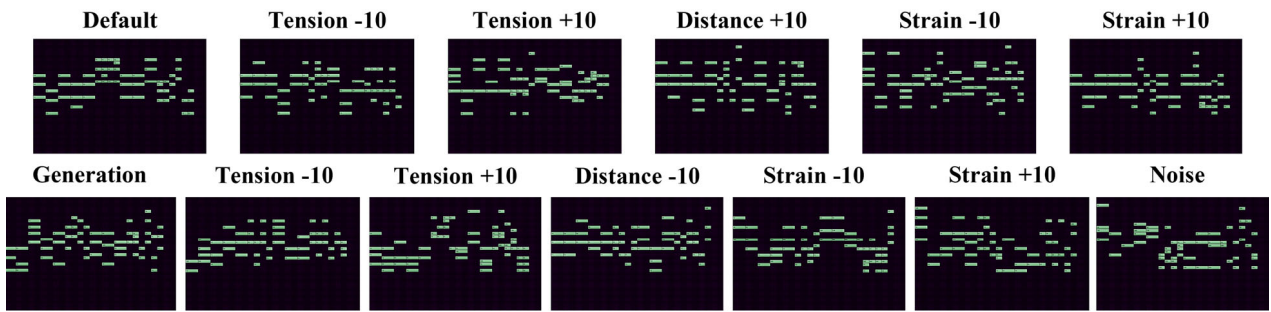

Figure A4: Visualization of reconstruction (first row) and generation (second row) results on a representative sample. 'Default' refers to the raw sample content before reconstruction. 'Generation' shows one representative generated chord progression. 'Noise' presents a chord progression recovered from randomly assigned perceptual feature sequence values.